\documentclass[runningheads]{llncs}
\usepackage{amsfonts}
\usepackage{epsfig}
\usepackage{amsmath}
\usepackage{graphicx}
\usepackage{amsmath}
\usepackage{amscd}
\usepackage{stmaryrd}
\usepackage{algorithm}
\usepackage{arydshln}
\usepackage{algorithmic}
\usepackage{youngtab}
\usepackage{young}
\usepackage{geometry}
\usepackage{bbm}
\usepackage{leftidx}
\usepackage{qcircuit}
\usepackage[colorlinks,
            linkcolor=blue,
            anchorcolor=blue,
            citecolor=blue,
            ]{hyperref}

\def\bt{\begin{theorem}}
\def\et{\end{theorem}}
\def\bp{\begin{proposition}}
\def\ep{\end{proposition}}
\def\bc{\begin{corollary}}
\def\ec{\end{corollary}}
\def\bo{\begin{proof}}
\def\eo{\end{proof}}
\def\bx{\begin{example}}
\def\ex{\end{example}}
\def\br{\begin{remark}}
\def\er{\end{remark}}
\def\bl{\begin{lemma}}
\def\el{\end{lemma}}

\def\bn{\begin{definition}}
\def\en{\end{definition}}
\def\ba{\begin{array}}
\def\ea{\end{array}}
\def\be{\begin{equation}}
\def\ee{\end{equation}}
\def\bd{\begin{description}}
\def\ed{\end{description}}
\def\bu{\begin{enumerate}}
\def\eu{\end{enumerate}}
\def\bi{\begin{itemize}}
\def\ei{\end{itemize}}

\newbox\bigstrutbox
\setbox\bigstrutbox=\hbox{\vrule height12.5pt depth5pt width0pt}
\def\bigstrut{\relax\ifmmode\copy\bigstrutbox\else\unhcopy\bigstrutbox\fi}

\newbox\Bigstrutbox
\setbox\Bigstrutbox=\hbox{\vrule height17.5pt depth5pt width0pt}
\def\Bigstrut{\relax\ifmmode\copy\Bigstrutbox\else\unhcopy\Bigstrutbox\fi}

\def\ds{\displaystyle}

\def\i{{\bf i}}

\def\u{{\bf u}}
\def\v{{\bf v}}

\def\x{{\bf x}}
\def\y{{\bf y}}
\def\z{{\bf z}}

\def\0{{\bf 0}}
\def\1{{\bf 1}}
\def\2{{\bf 2}}
\def\3{{\bf 3}}
\def\4{{\bf 4}}
\def\5{{\bf 5}}
\def\6{{\bf 6}}
\def\7{{\bf 7}}
\def\8{{\bf 8}}
\def\9{{\bf 9}}

\begin{document}

\pagestyle{headings}

\mainmatter

\title{Generalization of Quantum Fourier Transformation}

\titlerunning{Generalization of Quantum Fourier Transformation}

\author{Changpeng Shao \\
cpshao@amss.ac.cn}
\authorrunning{Changpeng Shao}

\institute{Academy of Mathematics and Systems Science, Chinese Academy of Sciences, Beijing 100190, China}

\maketitle

\begin{abstract}
  Quantum Fourier transformation is important in many quantum algorithms.
  In this paper, we generalize quantum Fourier transformation over the Abelian group $\mathbb{Z}_N$ from two different points to get more efficient unitary transformations.
  The obtained unitary transformations are given in concise and explicit formula which can be used directly.
  A relationship between the generalized quantum Fourier transformation and the dihedral hidden subgroup problem are discussed in this paper.
  This may lead a way to solve the dihedral hidden subgroup problem.
  The second goal of this paper is to give an explicit formula of quantum Haar transformation.
\end{abstract}

{\bf Key words: Quantum Fourier transformation, dihedral hidden subgroup problem, quantum Haar transformation}

\section{Introduction}
\setcounter{equation}{0}

Quantum computation only allows unitary transformations. However, there are only a few efficiently implemented unitary transformations we can use in the construction of quantum algorithm, such as Fourier transformation, Hadamard transformation, the reflection introduced in Grover's algorithm \cite{grover}, the exponential of sparse Hamiltonian used in linear system \cite{harrow} and some others. All these unitary transformations play important roles in quantum algorithm construction. For example, quantum Fourier transformation plays an important role in many quantum algorithms, most of them achieve an exponential speedup, such as quantum counting \cite{brassard}, polynomial interpolation \cite{childs2}, Pell's equation solving \cite{hallgren}, solving linear system \cite{harrow}, hidden subgroup problem \cite{kitaev}, factoring and discrete logarithm problem \cite{shor}, and so on.

There are also some works aims at finding new efficient unitary transformations. For example, in 1997, H{\o}yer \cite{hoyer} has constructed many efficiently implemented quantum transformations by generalized Kronecker product of matrices, including the quantum Haar transformation. In 2006, Bacon et.al. \cite{bacon} have considered the quantum circuits of Schur transform together with a clear interpretation in the generalized quantum phase estimation algorithm. Compared with the construction of new unitary transformations, another much more important problem is how to use these efficient unitary transformations to construct new quantum algorithms? However, this problem seems more difficult than constructing new unitary transformations. Anyway, as the construction tool of quantum algorithm, the finding of new efficient unitary transformations with explicit formula or with clear explanation is very important.

An important unitary transformation in quantum computing is Hadamard transformation. As a generalization of real Hadamard transformation, complex Hadamard transformation plays an important role in quantum information \cite{tadej}, \cite{werner} and quantum tomography \cite{wootters}. Quantum Fourier transformation belongs to the class of complex Hadamard transformation. Complex Hadamard transformation can also be used to construct unitary error bases \cite{klappenecker}, \cite{werner}, which are instrumental for quantum error-correcting codes and the design of teleportation and super-dense coding schemes. In the classical sense, it is an important task to construct inequivalent complex Hadamard transformation \cite{goyeneche}, \cite{haagerup}.

In this paper, we do some generalizations about quantum Fourier transformation.
Such a generalization is still in the Abelian group $\mathbb{Z}_N$, and mainly based on the circuit of quantum Fourier transformation.
The circuit of quantum Fourier transformation contains special properties like the input variables stay in triangular form.
By making the circuit more general, we can get different types of unitary transformations which contain concise and explicit expressions.
Also it contains Fourier transformation and Hadamard transformation as special cases. And it belongs to the class of complex Hadamard transformation in dimension $2^n$ for some $n$.

In order to find applications of the generalized Fourier transformation, we have consider the dihedral hidden subgroup problem which is related to the shortest vector problem \cite{regev}.
This generalized Fourier transformation seems more suitable to study dihedral hidden subgroup problem than Fourier transformation. Some ideas to solve the dihedral hidden subgroup problem are discussed in this paper. Since the shortest vector problem is NP-complete, so it may not easy to solve this problem in quantum computer. However, the generalized Fourier transformation might be a way to solve the dihedral hidden subgroup problem, and so solve the shortest vector problem.

The second target of this paper is to present an explicit formula about quantum Haar transformation. In \cite{hoyer}, a decomposition of quantum Haar transformation was obtained. In order to use a unitary transformation to construct quantum algorithms, it is much better and convenient to give its explicit expression. A final remark is that all the unitary transformations constructed in this paper can be efficiently implemented in quantum computer.

The structure of this paper is as follows: In section 2, we briefly introduce quantum Fourier transformation and its circuit. In section 3 and 4, we give two different types of generalization of quantum Fourier transformation. Also in section 3, the relation between the first type of generalized quantum Fourier transformation and the dihedral hidden subgroup problem are discussed. Finally, in section 5, the explicit formula of quantum Haar transformation is given.

\section{Quantum Fourier transformation}
\setcounter{equation}{0}

In this section, we consider a class of unitary transformations that can be efficiently implemented in quantum computer by generalizing quantum Fourier transformation from its implementation circuit.
Assume that $N=2^n$ for some $n$, denote $\omega_N=e^{2\pi \i/N}$. Then the Fourier transformation over $\mathbb{Z}_N$ is defined as:
\be\ba{rll}
F_N:\mathbb{Z}_N &\longrightarrow& \mathbb{Z}_N \\
|x\rangle        &\longmapsto& \ds\frac{1}{\sqrt{N}}\sum_{y=0}^{N-1}\omega_N^{xy}|y\rangle.
\ea\ee

By seting $x=x_0+x_12+\cdots+x_{n-1}2^{n-1},y=y_0+y_12+\cdots+y_{n-1}2^{n-1}$ and using the identity $\omega_N^N=1$, then we get the classical decomposition of Fourier transformation \cite{kaye}, \cite{nielsen}:
\[\ba{lll} \vspace{.2cm}
& & \ds\frac{1}{\sqrt{N}}\sum_{y=0}^{N-1}\omega_N^{xy}|y\rangle \\ \vspace{.2cm}
&=& \ds\frac{1}{\sqrt{N}}\sum_{y_0,y_1,\ldots,y_{n-1}=0}^1 \hspace{-.2cm} \omega_N^{2^{n-1}x_0y_{n-1}+2^{n-2}(x_0+2x_1)y_{n-2}+\cdots+(x_0+x_12+\cdots+x_{n-1}2^{n-1})y_0}|y_0,y_1,\cdots,y_{n-1}\rangle \\ \vspace{.2cm}
\ea\]
\[\ba{lll}
&=& \ds\frac{1}{\sqrt{N}}\Bigg(\sum_{y_{n-1}=0}^1 \omega_N^{2^{n-1}x_0y_{n-1}}|y_{n-1}\rangle\Bigg)\otimes\Bigg(\sum_{y_{n-2}=0}^1\omega_N^{2^{n-2}(x_0+2x_1)y_{n-2}}|y_{n-2}\rangle\Bigg)
    \otimes\cdots \\
& & \hfill \cdots\otimes\Bigg(\ds\sum_{y_0=0}^1\omega_N^{(x_0+x_12+\cdots+x_{n-1}2^{n-1})y_0}|y_0\rangle\Bigg)~ \\
&=& \ds\frac{1}{\sqrt{N}}\left(|0\rangle+\omega_N^{2^{n-1}x_0}|1\rangle\right)\otimes\left(|0\rangle+\omega_N^{2^{n-2}(x_0+2x_1)}|1\rangle\right)\otimes\cdots\otimes
    \left(|0\rangle+\omega_N^{x_0+x_12+\cdots+x_{n-1}2^{n-1}}|1\rangle\right).
\ea\]
So the circuit of Fourier transformation $F_N$ modulo some swap operations is

\[\hspace{-5.9cm}\Qcircuit @C=.7em @R=1.3em {
\lstick{|x_0\rangle}& \qw& \qw           & \ctrl{2}      & \qw & \ctrl{1}      & \gate{H}& \rstick{\ds\frac{1}{\sqrt{2}}\left(|0\rangle+\omega_N^{2^{n-1}x_0}|1\rangle\right) =: |u_0\rangle}\qw \\
\lstick{|x_1\rangle}& \qw& \ctrl{1}      & \qw           &\gate{H}& \gate{R_{n-2}}&  \qw& \rstick{\ds\frac{1}{\sqrt{2}}\left(|0\rangle+\omega_N^{2^{n-2}(x_0+2x_1)}|1\rangle\right)=: |u_1\rangle}\qw \\
\lstick{|x_2\rangle}& \gate{H} &\gate{R_{n-2}}& \gate{R_{n-3}}& \qw      &\qw &  \qw&\rstick{\ds\frac{1}{\sqrt{2}}\left(|0\rangle+\omega_N^{2^{n-3}(x_0+2x_1+2^2x_2)}|1\rangle\right)=: |u_2\rangle}\qw \\
&   &   & \lstick{\cdots\cdots} &  &  \\
\lstick{|x_{n-1}\rangle} &\qw &\qw & \lstick{\cdots\cdots}&\qw&\qw& \qw &\rstick{\ds\frac{1}{\sqrt{2}}\left(|0\rangle+\omega_N^{x_0+x_12+\cdots+x_{n-1}2^{n-1}}|1\rangle\right)=: |u_{n-1}\rangle}\qw \\
&&&&& \\
&&&& &&\lstick{\textmd{\textbf{Figure}~1}}& \\
&&&&& \\
}\]
where $R_i=\left(
             \begin{array}{cc} \vspace{.2cm}
               1 &~~ 0 \\
               0 &~~ \omega_N^{2^i} \\
             \end{array}
           \right)$ and $H=\ds\frac{1}{\sqrt{2}}\left(
             \begin{array}{rr} \vspace{.2cm}
               1 &~~ 1 \\
               1 &~~ -1\\
             \end{array}
           \right)$ is the Hadamard transformation.

In the above graph, we can find two basic properties about the relative phase in $|u_i\rangle$ with respect to the input variables $x_0,x_1,\ldots,x_{n-1}$:
\begin{enumerate}
  \item the input variables $x_0,x_1,\ldots,x_{n-1}$ stay in triangular form; \vspace{.1cm}
  \item the input variables $x_0,x_1,\ldots,x_{n-1}$ are independent to each other in each phase.
\end{enumerate}
By fixing the first properties, we can make some changes in the relative phase and obtain at least two different types of generalizations.

\section{First type of generalization of quantum Fourier transformation}
\setcounter{equation}{0}

By keeping the triangular form, we can make the relative phase more general. For $0\leq i\leq n-1$, setting
\be
|u_i\rangle=\frac{1}{\sqrt{2}}\left(|0\rangle+\omega_N^{\phi_{i0}x_0+\phi_{i1}x_1+\cdots+\phi_{ii}x_i}|1\rangle\right).
\ee

(1). $\ds|u_0\rangle=\frac{1}{\sqrt{2}}\left(|0\rangle+\omega_N^{\phi_{00}x_0}|1\rangle\right).$
Then $U=\ds\frac{1}{\sqrt{2}}\left(
                          \begin{array}{cc} \vspace{.2cm}
                            1 &~~ 1 \\
                            1 &~~ \omega_N^{\phi_{00}} \\
                          \end{array}
                        \right)$ is a unitary matrix. And this only happens in the case $\phi_{00}=2^{n-1}$, i.e., $U=H$ is the $2\times 2$ Hadamard transformation.

(2). $\ds|u_1\rangle=\frac{1}{\sqrt{2}}\left(|0\rangle+\omega_N^{\phi_{10}x_0+\phi_{11}x_1}|1\rangle\right).$ The phase $\omega_N^{\phi_{10}x_0}$ is obtained by a control transformation based on the first qubit $|x_0\rangle$, denote this phase transformation as $T(\phi_{10})$. The phase $\omega_N^{\phi_{11}x_1}$ is obtained from $|x_1\rangle$ directly. So $|u_1\rangle$ is obtained from the composition of $U_1$ and $T(\phi_{10})^{x_0}$ for some $U_1$:
\[|x_1\rangle\xrightarrow[]{U_1}\frac{1}{\sqrt{2}}\left(|0\rangle+\omega_N^{\phi_{11}x_1}|1\rangle\right)\xrightarrow[]{T(\phi_{10})^{x_0}}
\frac{1}{\sqrt{2}}\left(|0\rangle+\omega_N^{\phi_{10}x_0+\phi_{11}x_1}|1\rangle\right).\]
A similar analysis as to $|u_0\rangle$ shows that $U_1=H$ is the Hadamard transformation, so $\phi_{11}=2^{n-1}$. The following is the circuit of these two steps

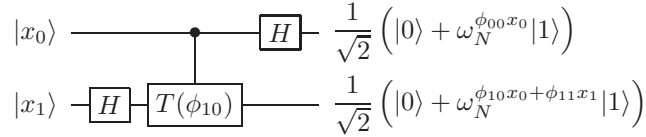
\[\hspace{-2.9cm}\Qcircuit @C=.7em @R=1.3em {
\lstick{|x_0\rangle}&\qw       & \ctrl{1}            & \gate{H} & \rstick{\ds\frac{1}{\sqrt{2}}\left(|0\rangle+\omega_N^{\phi_{00}x_0}|1\rangle\right)}\qw \\
\lstick{|x_1\rangle}& \gate{H} & \gate{T(\phi_{10})} & \qw      & \rstick{\ds\frac{1}{\sqrt{2}}\left(|0\rangle+\omega_N^{\phi_{10}x_0+\phi_{11}x_1}|1\rangle\right)}\qw
}\]

(3). In $|u_i\rangle$, define
\[\Phi_i=\phi_{i0}x_0+\phi_{i1}x_1+\cdots+\phi_{ii}x_i,\]
then $\ds|u_i\rangle=\frac{1}{\sqrt{2}}\left(|0\rangle+\omega_N^{\Phi_i}|1\rangle\right)$. Similar analysis shows that $\phi_{ii}=2^{n-1}$ and the phase $\phi_{ij}$ obtained from a control transformation based on the qubit $|x_j\rangle$ for $0\leq j\leq i-1$. This is just like the circuit implementation of Fourier transformation but changes $R_k$ into a general phase transformation.

Since
\[|u_0\rangle\otimes|u_1\rangle\otimes\cdots\otimes|u_{n-1}\rangle
=\frac{1}{\sqrt{N}}\sum_{y_0,\ldots,y_{n-1}=0}^1 \omega_N^{\Phi_0y_0+\Phi_1y_1+\cdots+\Phi_{n-1}y_{n-1}}|y_0,y_1,\cdots,y_{n-1}\rangle,\]
and
\[\Phi_0y_0+\Phi_1y_1+\cdots+\Phi_{n-1}y_{n-1}=\sum_{i=0}^{n-1}\sum_{j=0}^i\phi_{ij}x_jy_i.\]
So if denote the matrix
\be \label{matrix}
\Phi=\left(
       \begin{array}{cccc} \vspace{.1cm}
         \phi_{00}     ~~& \phi_{01}     ~~& \cdots ~~& \phi_{0(n-1)} \\ \vspace{.1cm}
         \phi_{10}     ~~& \phi_{11}     ~~& \cdots ~~& \phi_{1(n-1)} \\ \vspace{.1cm}
         \vdots        ~~& \vdots        ~~& \ddots ~~& \vdots \\
         \phi_{(n-1)0} ~~& \phi_{(n-1)1} ~~& \cdots ~~& \phi_{(n-1)(n-1)} \\
       \end{array}
     \right),
\ee
where
\be  \label{condition}
\left\{
  \begin{array}{ll} \vspace{.2cm}
    \phi_{ii}=2^{n-1}~\textmd{for}~0\leq i\leq n-1, & \hbox{} \\ \vspace{.2cm}
    \phi_{ij}~\textmd{is~an~integer~which~contains}~2^n~\textmd{as~a~factor~for}~i<j, & \hbox{} \\
    \phi_{ij}~\textmd{can~be~any~real~numbers~for}~i>j. & \hbox{}
  \end{array}
\right.
\ee
Then we have

\bt
Let $\Phi$ be the matrix in the form (\ref{matrix}) satisfies condition (\ref{condition}). Then the generalized Fourier transformation (also known as complex Hadamard transformation) from $\mathbb{Z}_N$ to itself is
\be \label{generalized Fourier transformation}
G_N|\x\rangle=\frac{1}{\sqrt{N}}\sum_{\y=0}^{N-1} \omega_N^{\y \Phi \x^T} |y\rangle,
\ee
where $\x=(x_0,\ldots,x_{n-1}), \y=(y_0,\ldots,y_{n-1})$.
Moreover, the circuit implementation complexity of $G_N$ is $O((\log N)^2)$.
\et

\bx
(1). $N=2,\Phi=1$, then formula (\ref{generalized Fourier transformation}) gives $2\times 2$ Hadamard transformation.

(2). $N=4,\Phi=\left(
                 \begin{array}{cc} \vspace{.1cm}
                   2 &~~ 0 \\
                   a &~~ 2 \\
                 \end{array}
               \right)$, then formula (\ref{generalized Fourier transformation}) gives
\[\frac{1}{2}\left(
               \begin{array}{rrrr} \vspace{.1cm}
                 1 &~~ 1 &~~ 1 &~~ 1 \\ \vspace{.1cm}
                 1 &~~ -1 &~~ 1 &~~ -1 \\ \vspace{.1cm}
                 1 &~~ \omega_4^a &~~ -1 &~~ -\omega_4^a \\
                 1 &~~ -\omega_4^a &~~ -1 &~~ \omega_4^a \\
               \end{array}
             \right).\]
\ex

The following are two special cases of (\ref{generalized Fourier transformation}):

(1). If $\Phi=2^{n-1}I_{n\times n}$, then $G_N=H^{\otimes n}$ is the Hadamard transformation.

(2). If $\Phi$ is the following Toeplitz matrix
\[\Phi=\left(
       \begin{array}{ccccc} \vspace{.1cm}
         2^{n-1} ~~& 2^{n}    ~~& \cdots ~~& 2^{2n-3} ~~& 2^{2n-2}  \\ \vspace{.1cm}
         2^{n-2} ~~& 2^{n-1}  ~~& \cdots ~~& 2^{2n-4} ~~& 2^{2n-3}  \\ \vspace{.1cm}
         \vdots  ~~& \vdots   ~~& \ddots ~~& \vdots   ~~& \vdots \\ \vspace{.1cm}
         2       ~~& 2^{2}    ~~& \cdots ~~& 2^{n-1}  ~~& 2^{n}  \\
         1       ~~& 2        ~~& \cdots ~~& 2^{n-2}  ~~& 2^{n-1}  \\
       \end{array}
     \right),\]
then $G_N$ is the composition of $O(n^2)$ swap transformations with Fourier transformation, since the right side of (\ref{generalized Fourier transformation}) is $\ds\frac{1}{\sqrt{N}}\sum_{y_0,y_1,\ldots,y_{n-1}=0}^1 \omega_N^{xy} |y_{n-1},\ldots,y_1,y_0\rangle.$

\br
In the above construction,

(1). We can change $\phi_{ij}x_j$ into a function $\phi_{ij}(x_j)$ of $x_j$ for $i>j$.

(2). We can also change $\phi_{i0}x_0+\cdots+\phi_{i(i-1)}x_{i-1}$ into a function $\phi_i(x_0,\ldots,x_{i-1})$ of $x_0,\ldots,x_{i-1}$ for $i>j$. In order to keep the circuit implementation complexity in $O(n^k)$ for some fixed $k$, the support of the function $\phi_i$ should be in size $O(n^k)$.
\er

\subsection{Further generalization of the transformation (\ref{generalized Fourier transformation})}

In (\ref{generalized Fourier transformation}), the matrix $\Phi$ should satisfy the condition (\ref{condition}). In this subsection, we try to find more relaxed condition of the matrix $\Phi$ such that (\ref{generalized Fourier transformation}) is a unitary transformation. At this time, the implementing efficiency may be affected.

Denote $T=\ds\frac{1}{\sqrt{N}}\sum_{\x,\y=0}^{N-1}\omega_N^{\y \Phi \x^T}|\y\rangle\langle\x|$, then
\[
TT^\dagger=\frac{1}{N}\sum_{x,y,u,v=0}^{N-1}\omega_N^{\y \Phi \x^T-\u \Phi \v^T}|\y\rangle\langle\x|\v\rangle\langle\u|
=\ds\frac{1}{N}\sum_{y,u=0}^{N-1}\left(\sum_{\x=0}^{N-1}\omega_N^{(\y-\u) \Phi \x^T}\right)|\y\rangle\langle\u|.
\]
For any $\z\in\mathbb{Z}_N$, define
\be
A(\z):=\frac{1}{N}\sum_{\x=0}^{N-1}\omega_N^{\z\Phi \x^T},
\ee
then $T$ is unitary if and only if $A(\z)=\delta^0_\z$.

On the other hand, if denote $\tilde{\z}=\z\Phi=(\tilde{z}_0,\ldots,\tilde{z}_{n-1})$, then
\be
A(\z)=\frac{1}{N}\sum_{x_0,\ldots,x_{n-1}=0}^1\omega_N^{x_0\tilde{z}_0+\cdots+x_{n-1}\tilde{z}_{n-1}}
=\frac{1}{N}\prod_{j=0}^{n-1}(1+\omega_N^{\tilde{z}_j}).
\ee
So
\begin{enumerate}
  \item $A(\z)=0$ if and only if there exists $0\leq j\leq n-1$, such that $\tilde{z}_j\equiv 2^{n-1}\mod N$.\vspace{.1cm}
  \item $A(\z)=1$ if and only if $\tilde{\z}=0$.
\end{enumerate}
Note that $\z\in\mathbb{Z}_2^n$, so $\tilde{z}_j$ equals the summation of some elements of $\Phi$ in the $j$-th column. Since in order to make sure $T$ is unitary, $A(\z)=0$ for all $\z\neq 0$. This means for any $1\leq r\leq n$ and $1\leq i_1<i_2<\cdots<i_r \leq n$, there is a $1\leq j\leq n$, such that the summation of the $r$ elements in the $i_1,i_2,\ldots,i_r$-th row and in the $j$-th column equals $2^{n-1}\mod N$. Therefore, we have

\bp\label{generalized Fourier transformation:more}
Let $\Phi$ be any $n\times n$ matrix, then $T=\ds\frac{1}{\sqrt{N}}\sum_{\x,\y=0}^{N-1}\omega_N^{\y \Phi \x^T}|\y\rangle\langle\x|$ is unitary if and only if for any $1\leq r\leq n$ and $1\leq i_1<i_2<\cdots<i_r \leq n$, there is a $1\leq j\leq n$, such that the summation of the $r$ elements in the $i_1,i_2,\ldots,i_r$-th row and in the $j$-th column equals $2^{n-1}\mod N$.
\ep

For example, in the case $n=2$, the matrix $\Phi$ can be $\left(
                 \begin{array}{cc} \vspace{.1cm}
                   2 &~~ 0 \\
                   a &~~ 2 \\
                 \end{array}
               \right),\left(
                 \begin{array}{cc} \vspace{.1cm}
                   2 &~~ a \\
                   0 &~~ 2 \\
                 \end{array}
               \right)$ and $\left(
                 \begin{array}{cc} \vspace{.1cm}
                   2 &~~ a \\
                   2 &~~ 2-a \\
                 \end{array}
               \right)$, where $a\in\mathbb{R}$.
Also, if the matrix in (\ref{matrix}) satisfies condition (\ref{condition}), then transpose any row into its corresponding column will give a matrix satisfies the condition discussed in proposition \ref{generalized Fourier transformation:more}.
So the transformation considered in proposition \ref{generalized Fourier transformation:more} can have more structures than (\ref{generalized Fourier transformation}).

\subsection{The relation of the transformation (\ref{generalized Fourier transformation}) to dihedral hidden subgroup problem}

In this subsection, we give an application of the transformation (\ref{generalized Fourier transformation}) to dihedral hidden subgroup problem. This may not solve the dihedral hidden subgroup problem very well, but it provides an ideal to solve it.
Consider the following procedure, where $z$ is a fixed vector of length $n$, whose entries can be any real number:
\be \label{procedure2}
\frac{1}{\sqrt{N}}\sum_{x=0}^{N-1}\omega_N^{z x^T}|x\rangle\rightarrow\frac{1}{N}\sum_{y=0}^{N-1}\left(\sum_{x=0}^{N-1}\omega_N^{z x^T-y\Phi x^T}\right)|y\rangle.
\ee
Denote~$z=(z_0,z_1,\ldots,z_{n-1}),\tilde{y}=yM=(\tilde{y}_0,\tilde{y}_1,\ldots,\tilde{y}_{n-1})$, then the probability of $|y\rangle$ is
\be
p:=\frac{1}{N^2}\left|\sum_{x=0}^{N-1}\omega_N^{(z-y\Phi)x^T}\right|^2=\prod_{i=0}^{n-1}\cos^2\left(\frac{z_i-\tilde{y}_i}{N}\pi\right).
\ee
The ideal case is $z=y\Phi\mod N$. 

Recall the dihedral hidden subgroup problem considered in \cite{kuperberg}, \cite{regev}. Given several states in the form $\ds\frac{1}{\sqrt{2}}(|0\rangle+\omega_N^{ds}|1\rangle)$, where $s$ is known and $d$ is unknown. The problem is from these states to compute $d$. In (\ref{procedure2}), if choose $z=(ds_0,ds_1,\ldots,ds_{n-1})$, denote $N=2^n$, then we have
\be\ba{lll} \vspace{.2cm}
|\psi\rangle &=& \ds\frac{1}{\sqrt{N}}(|0\rangle+\omega_N^{ds_0}|1\rangle)\otimes(|0\rangle+\omega_N^{ds_1}|1\rangle)\otimes\cdots\otimes(|0\rangle+\omega_N^{ds_{n-1}}|1\rangle) \\ \vspace{.2cm}
             &=& \ds\frac{1}{\sqrt{N}}\sum_{x=0}^{N-1}\omega_N^{zx^T}|x\rangle \\
             &\rightarrow& \ds\frac{1}{N}\sum_{y=0}^{N-1}\left(\sum_{x=0}^{N-1}\omega_N^{(z-y\Phi) x^T}\right)|y\rangle.
\ea\ee

Set
\[\ba{lll} \vspace{.2cm}
d   &=& d_0+d_12+\cdots+d_{n-1}2^{n-1}, \\
s_i &=& s_{i0}+s_{i1}2+\cdots+s_{i,n-1}2^{n-1},~(i=0,1,\ldots,n-1).
\ea\]
Then
\[ds_i = S_{0i}d_0+S_{1i}d_1+\cdots+S_{n-2,i}d_{n-2}+S_{n-1,i}d_{n-1}\mod 2^n,\]
where
\[S_{ji}=(s_{i0}+s_{i1}2+\cdots+s_{i,n-j-1}2^{n-j-1})2^j=s_i2^j \mod 2^n,~(j=0,1,\ldots,n-1).\]

We choose the matrix $\Phi=(\phi_{ij})$ in the down triangular form, such that
\be\ba{lll}\vspace{.2cm}
\phi_{ii} &=& 2^{n-1},\hspace{.67cm} (0\leq i\leq n-1); \\
\phi_{ji} &=& S_{n-j-1,i},~(0\leq i\leq n-2, i+1\leq j\leq n-1).
\ea\ee
So
\be
\lambda_i:=ds_i-\tilde{y}_i=S_{n-i-1,i}d_{n-i-1}+S_{n-i,i}d_{n-i}+\cdots+S_{n-1,i}d_{n-1}-2^{n-1}d_{n-i-1}.
\ee

In this construction, we should reasonable to choose $s_0,s_1,\ldots,s_{n-1}$ in some suitable order, such that we have a high probability to get~$d_0=y_{n-1},d_1=y_{n-2},\ldots,d_{n-1}=y_0$.

Since
\[\ba{lll} \vspace{.2cm}
\lambda_0 &=& S_{n-1,0}d_{n-1}-2^{n-1}d_{n-1} = 2(s_{00}-1)d_{n-1}, \\\vspace{.2cm}
\lambda_1 &=& S_{n-2,1}d_{n-2}+S_{n-1,1}d_{n-1}-2^{n-1}d_{n-2}=2^{n-2}s_{10}(d_{n-2}+2d_{n-1})+2^{n-1}(s_{11}-1)d_{n-2}, \\
\lambda_i &=& \ds\sum_{k=0}^{i-1}s_{ik}2^k\left(\sum_{j=0}^{i-k}d_{n-i+j-1}2^{n-i+j-1}\right)+2^{n-1}(s_{ii}-1)d_{n-i-1},~(i=0,1,\ldots,n-1).
\ea\]
For simplicity, denote $D_{ik}=\ds\sum_{j=0}^{i-k}d_{n-i+j-1}2^{n-i+j-1}$ and $\tilde{s}_{ik}=\left\{
                                                                                          \begin{array}{ll} \vspace{.1cm}
                                                                                            2^ks_{ik}, & \hbox{$k<i$;} \\
                                                                                            2^k(s_{ik}-1), & \hbox{$k=i$.}
                                                                                          \end{array}
                                                                                        \right.$ Then
\[\lambda_i=\ds\sum_{k=0}^{i}\tilde{s}_{ik}D_{ik}=(\tilde{s}_{i0},\tilde{s}_{i1},\ldots,\tilde{s}_{ii})\cdot(D_{i0},D_{i1},\ldots,D_{ii})=:\tilde{s}_i\cdot D_i.\]

Note that $d=d_0+d_12+\cdots+d_{n-i-2}2^{n-i-2}+D_{ik}+d_{n-k}2^{n-k}+\cdots+d_{n-1}2^{n-1}$.
This means $\tilde{s}_i$ is better to be orthogonal with $D_i$ module $2^n$, it has a freedom of $i-f_i+n-i-1=n-f_i-1$, where $f_i$ is the number of nonzero entry in $D_i$. Denote $f=\max\{f_0,f_1,\ldots,f_{n-1}\}$. So theoretically, we can find $s_i$ in $O(2^{f_i})$ time, and the total complexity to find good $s_0,s_1,\ldots,s_{n-1}$ is $O(2^{f_0}+2^{f_1}+\cdots+2^{f_{n-1}})=O(2^f)$. This is also the complexity to compute $d$.

As we can see, the best case is the matrix formed by the binary expanding of $s_0,s_1,\ldots,s_{n-1}$ is a upper triangular matrix with diagonal entries 1. In this case, $\lambda_i=0$ for all $i$, and so $p=1$, i.e., we can find $d$ deterministically. We should measure about $2^n$ times to such $s_0,s_1,\ldots,s_{n-1}$. The sieve method introduced by Kuperberg \cite{kuperberg} can find the state $\ds\frac{1}{\sqrt{2}}(|0\rangle+\omega_N^{d2^{n-1}}|1\rangle)=\frac{1}{\sqrt{2}}(|0\rangle+(-1)^d|1\rangle)$ in time $2^{O(\sqrt{n})}$. Within the same complexity, we can find $s_0,s_1,\ldots,s_{n-1}$ as we wanted. So this will compute $d$ in time $2^{O(\sqrt{n})}$. On the other hand, if we set $n=k+l$, such that we use the Kuperberg sieve method to find $s_0,s_1,\ldots,s_{k-1}$ with good property. The remaining $s_k,\ldots,s_{n-1}$ are choosing randomly. If we are lucky enough, then the probability $p=2^{-cl}$ for some constant $c$. The total complexity is $O(2^{\sqrt{k}}+2^{c\sqrt{l}})$. Then we should choose $\sqrt{k}=\sqrt{l}$, i.e., $k=l=n/2$. At this time, the complexity is still $2^{O(\sqrt{n})}$.

The choosing of $\Phi$ in the above process may not good enough. Another idea is using the more general form of (\ref{generalized Fourier transformation}), i.e., the transformation considered in proposition \ref{generalized Fourier transformation:more}.
Since $d=(d_0,d_1,\ldots,d_{n-1})\cdot (2^0,2^1,\ldots,2^{n-1})$, denote
\[\Phi_0=\left(
           \begin{array}{c}\vspace{.2cm}
             2^0 \\\vspace{.2cm}
             2^1 \\\vspace{.2cm}
             \vdots \\
             2^{n-1} \\
           \end{array}
         \right)(s_0,s_1,\ldots,s_{n-1})=\left(
                                           \begin{array}{cccc}\vspace{.2cm}
                                             s_0 &~~ s_1 &~~ \cdots &~~ s_{n-1} \\\vspace{.2cm}
                                             2s_0 &~~ 2s_1 &~~ \cdots &~~ 2s_{n-1} \\\vspace{.2cm}
                                             \vdots &~~ \vdots &~~ \ddots &~~ \vdots \\
                                             2^{n-1}s_0 &~~ 2^{n-1}s_1 &~~ \cdots &~~ 2^{n-1}s_{n-1} \\
                                           \end{array}
                                         \right).\]
Then
\[p:=\textmd{Prob}(|y\rangle=|d\rangle)=\frac{1}{N^2}\left|\sum_{x=0}^{N-1}\omega_N^{d(\Phi_0-\Phi)x^T}\right|^2,\]
where $\Phi$ is a $n\times n$ matrix satisfies the condition described in proposition \ref{generalized Fourier transformation:more}. It remains a problem to find suitable $\Phi$ to make $p$ large.

\section{Second type of generalization of quantum Fourier transformation}
\setcounter{equation}{0}

The above generalization only makes the relative phase more general. In this subsection, we still keep the triangular form, but instead we will make the control transformation more general, not just change the relative phase.
We will change $H$ and $R_i$ into any $2\times 2$ unitary transformation in Figure 1 as follows.

Let $C_0$ be a $2\times 2$ unitary transformation and for $0\leq i\leq n-2$, define the following control transformation based on the first $i+1$ quibits
\be
C_{i+1}=\sum_{j_0,j_1,\ldots,j_i=0}^1|j_0,j_1,\cdots,j_i\rangle\langle j_0,j_1,\cdots,j_i|\otimes T_{j_0,j_1,\ldots,j_i},
\ee
where $T_{j_0,j_1,\ldots,j_i}$ are $2\times 2$ unitary transformations.
It means, when $|x_0,x_1,\cdots,x_i\rangle=|j_0,j_1,\cdots,j_i\rangle$, then we apply $T_{j_0,j_1,\ldots,j_i}$ on $|x_{i+1}\rangle$. Let $\widetilde{G}_N$ be the unitary transformation obtained in Figure 1 first by changing the Hadamard transformation $H$ in the first line into $C_0$, and deleting all other Hadamard transformations in other lines. Then changing all the control transformation into $C_i,$ see the following circuit

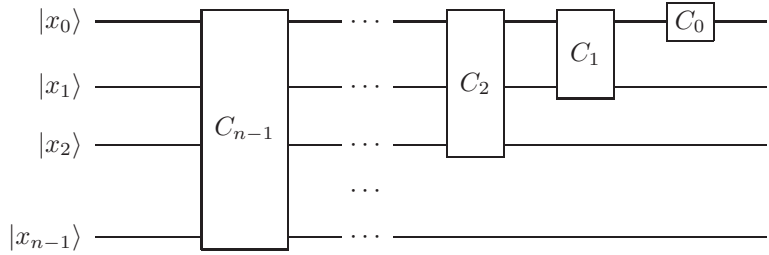
\[\Qcircuit @C=2em @R=1.3em {
\lstick{|x_0\rangle}&\qw      & \multigate{4}{C_{n-1}}     & \qw & \lstick{\cdots} & \multigate{2}{C_2}            & \multigate{1}{C_1}      & \gate{C_0} & \qw \\
\lstick{|x_1\rangle}&\qw      & \ghost{C_{n-1}}            & \qw & \lstick{\cdots} & \ghost{C_2}            & \ghost{C_1}    & \qw        & \qw \\
\lstick{|x_2\rangle}&\qw      & \ghost{C_{n-1}}            & \qw & \lstick{\cdots} & \ghost{C_2} & \qw                 & \qw        & \qw \\
 & & & &\lstick{\cdots} & & \\
\lstick{|x_{n-1}\rangle}&\qw  & \ghost{C_{n-1}} & \qw            & \lstick{\cdots} & \qw & \qw & \qw & \qw\\
&&&&& \\
}\]

For consistence, for any $k\in\{0,1\}$, we denote
\be\ba{rll}\vspace{.2cm}
C_0|k\rangle &=& t_0(k)|0\rangle+t_1(k)|1\rangle, \\
T_{j_0,j_1,\ldots,j_i}|k\rangle &=& t_0(j_0,j_1,\ldots,j_i,k)|0\rangle+t_1(j_0,j_1,\ldots,j_i,k)|1\rangle.
\ea\ee
So we have the following formula about $\widetilde{G}_N$
\be\ba{lll}\vspace{.2cm}
& & \widetilde{G}_N|x_0,x_1,\cdots,x_{n-1}\rangle \\\vspace{.2cm}
&=& \Big(t_0(x_0)|0\rangle+t_1(x_0)|1\rangle\Big)\otimes\Big(t_0(x_0,x_1)|0\rangle+t_1(x_0,x_1)|1\rangle\Big)\otimes\cdots \\\vspace{.2cm}
& & \hfill \cdots\otimes \Big(t_0(x_0,x_1,\ldots,x_{n-1})|0\rangle+t_1(x_0,x_1,\ldots,x_{n-1})|1\rangle\Big) \\
&=& \ds\sum_{y_0,y_1,\ldots,y_{n-1}=0}^1t_{y_0}(x_0)t_{y_1}(x_0,x_1)\cdots t_{y_{n-1}}(x_0,x_1,\ldots,x_{n-1})|y_0,y_1,\cdots,y_{n-1}\rangle.
\ea\ee

Next, we give two special cases about $\widetilde{G}_N$ based on rotations:
Since the notation in the above formula are too abstract, so in the following we construct $\widetilde{G}_N$ step by step.

(1). Let $C_0=H$, then $|x_0\rangle\mapsto \ds\frac{1}{\sqrt{2}}\Big(|0\rangle+(-1)^{x_0}|1\rangle\Big)$. For any $n-1\geq i>j\geq 0$ and $s\in\{0,1\}$, denote
\be\label{rotation}
R(\theta_{ij}(s))=\left(
                     \begin{array}{rr} \vspace{.2cm}
                        \cos[\theta_{ij}(s)]  &~~ \sin[\theta_{ij}(s)] \\
                        -\sin[\theta_{ij}(s)] &~~ \cos[\theta_{ij}(s)] \\
                     \end{array}
                  \right)
\ee
as a rotation depend on $s$ and $i,j$ with rotational angle $\theta_{ij}(s)$. Then $|x_1\rangle$ is changed into
\[\ba{lcl}\vspace{.2cm}
&& \ds\frac{1}{\sqrt{2}}\Big(|0\rangle+(-1)^{x_1}|1\rangle\Big) \xrightarrow[]{R(\theta_{10}(x_0))} \\\vspace{.2cm}
&& \ds\frac{1}{\sqrt{2}}\Big[\cos[\theta_{10}(x_0)]|0\rangle-\sin[\theta_{10}(x_0)]|1\rangle
+(-1)^{x_1}\Big(\sin[\theta_{10}(x_0)]|0\rangle+\cos[\theta_{10}(x_0)]|1\rangle\Big)\Big] \\
&=& \ds\frac{1}{\sqrt{2}}\Big[\Big(\cos[\theta_{10}(x_0)]+(-1)^{x_1}\sin[\theta_{10}(x_0)]\Big)|0\rangle+(-1)^{x_1}\Big(\cos[\theta_{10}(x_0)]-(-1)^{x_1}\sin[\theta_{10}(x_0)]\Big)|1\rangle\Big].
\ea\]
Similarly, we construct a control transformation that maps $|x_2\rangle$ into
\[\ba{lcl}\vspace{.2cm}
|x_2\rangle &\longrightarrow& \ds\frac{1}{\sqrt{2}}\Big(|0\rangle+(-1)^{x_2}|1\rangle\Big) \\\vspace{.2cm}
&\xrightarrow[]{R(\theta_{21}(x_1))R(\theta_{20}(x_0))}& \ds\frac{1}{\sqrt{2}}\Big[\cos[\theta_{20}(x_0)+\theta_{21}(x_1)]|0\rangle-\sin[\theta_{20}(x_0)+\theta_{21}(x_1)]|1\rangle \\\vspace{.2cm}
& & \hfill \ds+(-1)^{x_2}\Big(\sin[\theta_{20}(x_0)+\theta_{21}(x_1)]|0\rangle+\cos[\theta_{20}(x_0)+\theta_{21}(x_1)]|1\rangle\Big)\Big]~ \\\vspace{.2cm}
&=& \ds\frac{1}{\sqrt{2}}\Big[\Big(\cos[\theta_{20}(x_0)+\theta_{21}(x_1)]+(-1)^{x_2}\sin[\theta_{20}(x_0)+\theta_{21}(x_1)]\Big)|0\rangle \\
& & \hspace{.8cm}\ds+(-1)^{x_2}\Big(\cos[\theta_{20}(x_0)+\theta_{21}(x_1)]-(-1)^{x_2}\sin[\theta_{20}(x_0)+\theta_{21}(x_1)]\Big)|1\rangle\Big].
\ea\]
For simplicity, denote
\be \label{Theta}
\Theta_j=\theta_{j0}(x_0)+\theta_{j1}(x_1)+\cdots+\theta_{j(j-1)}(x_{j-1}),
\ee
then construct control transformation that maps $|x_j\rangle$ into
\[\ba{lll}\vspace{.2cm}
|x_j\rangle &\rightarrow& \ds\frac{1}{\sqrt{2}}\Big[\Big(\cos\Theta_j+(-1)^{x_j}\sin\Theta_j\Big)|0\rangle+(-1)^{x_j}\Big(\cos\Theta_j-(-1)^{x_j}\sin\Theta_j\Big)|1\rangle\Big] \\
            &=& \ds\frac{1}{\sqrt{2}}\sum_{y_j=0}^1(-1)^{x_jy_j}\Big(\cos\Theta_j+(-1)^{x_j+y_j}\sin\Theta_j\Big)|y_j\rangle
\ea\]
Therefore at this time $\widetilde{G}_N$ has the following formula
\be\ba{lll}\vspace{.2cm}\label{trans1}
& & \widetilde{G}_N|x_0,x_1,\cdots,x_{n-1}\rangle \\\vspace{.2cm}
&=& \ds\frac{1}{\sqrt{N}}\sum_{y_0,y_1,\ldots,y_{n-1}=0}^1(-1)^{x\cdot y}\prod_{j=1}^{n-1}\Bigg[\cos\Theta_j+(-1)^{x_j+y_j}\sin\Theta_j \Bigg]|y_0,y_1,\cdots,y_{n-1}\rangle.
\ea\ee

(2). In the above case, we use Hadamard transformation in the first step, now we change it into rotations. If $\alpha_j(0)$ is a rotational angle, then we set $\alpha_j(1)=\alpha_j(0)-\pi/2$. So any $2\times 2$ rotation with rotational angle $\alpha_j(0)$ has the following expression
\[|x_0\rangle\rightarrow \cos[\alpha_0(x_0)]|0\rangle-\sin[\alpha_0(x_0)]|1\rangle.\]
Use the nations in (\ref{rotation}), then construct a control transformation that maps $|x_1\rangle$ into
\[|x_1\rangle \rightarrow\cos[\alpha_1(x_1)]|0\rangle-\sin[\alpha_1(x_1)]|1\rangle
\xrightarrow[]{R(\theta_{10}(x_0))} \cos[\alpha_1(x_1)+\theta_{10}(x_0)]|0\rangle-\sin[\alpha_1(x_1)+\theta_{10}(x_0)]|1\rangle.\]
Similarly, construct control transformation satisfies
\[\ba{lll}\vspace{.2cm}
|x_2\rangle
&\rightarrow& \cos[\alpha_2(x_2)+\theta_{20}(x_0)+\theta_{21}(x_1)]|0\rangle-\sin[\alpha_2(x_2)+\theta_{20}(x_0)+\theta_{21}(x_1)]|1\rangle \\
&=& \ds \sum_{y_2=0}^1\cos\left[\alpha_2(x_2)+\theta_{20}(x_0)+\theta_{21}(x_1)+\frac{\pi y_2}{2}\right]|y_2\rangle.\ea\]

We can simplify denote
\be\ba{rll} \label{Psi}\vspace{.2cm}
\theta_{jj}(x_j) &=& \alpha_j(x_j), \\
\Psi_j &=& \theta_{j0}(x_0)+\theta_{j1}(x_1)+\cdots+\theta_{jj}(x_j).
\ea\ee
Then
\[|x_j\rangle\rightarrow\cos\Psi_j|0\rangle-\sin\Psi_j|1\rangle=\sum_{y_j=0}^1\cos\left[\Psi_j+\frac{\pi y_j}{2}\right]|y_j\rangle.\]
Therefore, in this construction

\be\ba{lll}\vspace{.2cm}\label{trans2}
& & \widetilde{G}_N|x_0,x_1,\cdots,x_{n-1}\rangle \\\vspace{.2cm}
&=& \ds\sum_{y_0,y_1,\ldots,y_{n-1}=0}^1 ~\prod_{j=0}^{n-1} \cos\left[\Psi_j+\frac{\pi y_j}{2}\right]
    |y_0,y_1,\cdots,y_{n-1}\rangle \\\vspace{.2cm}
&=& \ds\frac{1}{2}\sum_{y_0,y_1,\ldots,y_{n-1}=0}^1 ~\prod_{j=0}^{n-1}
    \Bigg(\textmd{exp}\left[\i\left(\Psi_j+\frac{\pi y_j}{2}\right)\right]+~\ds\textmd{exp}\left[-\i\left(\Psi_j+\frac{\pi y_j}{2}\right)\right]\Bigg)
    |y_0,y_1,\cdots,y_{n-1}\rangle \\
&=& \ds\frac{1}{2}\sum_{y_0,y_1,\ldots,y_{n-1}=0}^1~\sum_{k_0,k_1,\ldots,k_{n-1}=0}^1
    \textmd{exp}\left[\i\sum_{j=0}^{n-1}(-1)^{k_j}\left(\Psi_j+\frac{\pi y_j}{2}\right)\right]|y_0,y_1,\cdots,y_{n-1}\rangle.
\ea\ee

\bt
For any $0\leq j\leq i\leq n-1$, let $\theta_{ij}$ be a function from $\{0,1\}$ to $[0,2\pi)$ and $\theta_{ii}(1)=\theta_{ii}(0)-\pi/2$. Define $\Theta_j,\Phi_j$ as in formula (\ref{Theta}), (\ref{Psi}) for any $0\leq j\leq n-1$.
Then the unitary transformations defined by formula (\ref{trans1}) and (\ref{trans2}) are efficiently implemented in quantum computer with complexity $O(n^2)$.
\et

\br
Compared with (\ref{generalized Fourier transformation}), the expression of unitary transformations given in (\ref{trans1}) and (\ref{trans2}) are not so good. And it seems not so easy to find applications of them from their expressions.
\er

\section{Quantum Haar  transformation}
\setcounter{equation}{0}

In this section, we consider the quantum Haar transformation, which has been considered in \cite{fijany},\cite{hoyer}, but without an explicit formula of quantum Haar transformation.
In order to avoid confusing, we use $P_N$ to denote $N\times N$ Haar  transformation and $A_N$ to denote $N\times N$ Haar matrix. The difference is $P_N$ is unitary, but $A_N$ is not unitary. And $P_N$ is obtained by changing $A_N$ into unitary.

The $2\times 2$ Haar matrix is defined as:
\be \label{haar matrix: d=2}
A_2=\left(
        \begin{array}{rr}\vspace{.2cm}
          1 &~~ 1 \\
          1 &~~ -1 \\
        \end{array}
      \right).
\ee
The general $2^n\times 2^n$ Haar matrix can be derived by the following equation:
\be \label{haar matrix}
A_{2^n}=\left(
        \begin{array}{ll}\vspace{.2cm}
          A_{2^{n-1}}\otimes [1,1] \\
          I_{2^{n-1}}\otimes [1,-1] \\
        \end{array}
      \right),
\ee
where $[1,1]=\langle0|+\langle1|$ and $[1,-1]=\langle0|-\langle1|$ are row vectors.

Haar matrix will be changed into a unitary matrix by normalizing each row. The normalized Haar matrix is Haar  transformation $P_N$. For example,
\[P_2=\frac{1}{\sqrt{2}}\left(
        \begin{array}{rr}\vspace{.2cm}
          1 ~~& 1 \\
          1 ~~& -1 \\
        \end{array}
      \right), \hspace{.3cm} P_4=\frac{1}{2}\left(
        \begin{array}{rrrr}\vspace{.2cm}
          1 ~~& 1 ~~& 1 ~~& 1 \\\vspace{.2cm}
          1 ~~& 1 ~~& -1 ~~& -1 \\\vspace{.2cm}
          \sqrt{2} ~~& -\sqrt{2} ~~& 0 ~~& 0 \\
          0 ~~& 0 ~~& \sqrt{2} ~~& -\sqrt{2} \\
        \end{array}
      \right).\]
As we can see $P_2$ is Hadamard transformation $H$. Compared with Hadamard transformation whose entries are $+1$ and $-1$, there are many elements with zero value in the Haar  transformation, so the computation time is short.

Quantum Haar  transformation is a direct generalization of Haar  transformation to quantum operator, and it has been considered in the past \cite{hoyer}. In the following, we consider the explicit formula of quantum Haar  transformation operating on basis.

\bl
About the Haar matrix (\ref{haar matrix}), we have the following identity
\be\label{haar matrix:identity}
A_{2^n}|x_0,\ldots,x_{n-1}\rangle=|0\rangle^{\otimes n}+\sum_{i=0}^{n-1}(-1)^{x_i}|0\rangle^{\otimes (n-i-1)}|1\rangle|x_0,\ldots,x_{i-1}\rangle.
\ee

\el

\bo We prove (\ref{haar matrix:identity}) by induction. If $n=1$, then from (\ref{haar matrix:identity}), we have $A_2|x_0\rangle=|0\rangle+(-1)^{x_0}|1\rangle$, which gives the same matrix as (\ref{haar matrix: d=2}). So (\ref{haar matrix:identity}) holds for $n=1$. Assume that (\ref{haar matrix:identity}) holds for $n=m-1$, then we need prove it holds for $n=m$.

For simplicity, denote $|\tilde{x}\rangle=|x_0,\ldots,x_{m-2}\rangle$, so $|x\rangle=|\tilde{x}\rangle\otimes |x_{m-1}\rangle$. Then by definition
\[A_{2^m}|x\rangle=\left(
        \begin{array}{ll}\vspace{.2cm}
          A_{2^{m-1}}\otimes [1,1] \\
          I_{2^{m-1}}\otimes [1,-1] \\
        \end{array}
      \right)|x\rangle=\left(
        \begin{array}{ll}\vspace{.2cm}
          A_{2^{m-1}}|\tilde{x}\rangle\otimes [1,1]|x_{m-1}\rangle \\
          I_{2^{m-1}}|\tilde{x}\rangle\otimes [1,-1]|x_{m-1}\rangle \\
        \end{array}
      \right).\]
By introduction,
\[A_{2^{m-1}}|\tilde{x}\rangle=|0\rangle^{\otimes (m-1)}+\sum_{i=0}^{m-2}(-1)^{x_i}|0\rangle^{\otimes (m-i-2)}|1\rangle|x_0,\ldots,x_{i-1}\rangle.\]

On one hand, $[1,1]|x_{m-1}\rangle=\langle 0|x_{m-1}\rangle+\langle 1|x_{m-1}\rangle=1$, so
\be \label{up}
\left(
        \begin{array}{cc}\vspace{.2cm}
          A_{2^{m-1}}|\tilde{x}\rangle\otimes [1,1]|x_{m-1}\rangle \\
          0 \\
        \end{array}
      \right)=|0\rangle A_{2^{m-1}}|\tilde{x}\rangle=|0\rangle^{\otimes m}+\sum_{i=0}^{m-2}(-1)^{x_i}|0\rangle^{\otimes (m-i-1)}|1\rangle|x_0,\ldots,x_{i-1}\rangle.
\ee
On the other hand, $[1,-1]|x_{m-1}\rangle=\langle 0|x_{m-1}\rangle-\langle 1|x_{m-1}\rangle=(-1)^{x_{m-1}}$, so
\be \label{down}
\left(
        \begin{array}{cc}\vspace{.2cm}
          0 \\
          I_{2^{m-1}}|\tilde{x}\rangle\otimes [1,-1]|x_{m-1}\rangle \\
        \end{array}
      \right)=(-1)^{x_{m-1}}|1\rangle I_{2^{m-1}}|\tilde{x}\rangle=(-1)^{x_{m-1}}|1\rangle|\tilde{x}\rangle.
\ee
Finally, adding the two results (\ref{up}), (\ref{down}), we get the formula (\ref{haar matrix:identity}) for $n=m$.
\hfill $\square$
\eo

\bt
The Quantum Haar  transformation has the following expression by applying it on the basis
\be\label{haar}
P_{2^n}:|x_0,\ldots,x_{n-1}\rangle\mapsto \ds\frac{1}{\sqrt{2^n}}\Big(|0\rangle^{\otimes n}+\sum_{i=0}^{n-1}(-1)^{x_i}\sqrt{2^i}|0\rangle^{\otimes (n-i-1)}|1\rangle|x_0,\ldots,x_{i-1}\rangle\Big).
\ee
Moreover, the inverse of Haar  transformation satisfies, for any $0\leq i\leq n-1$:
\be\label{haar:inverse}
P_{2^n}^\dag:|0\rangle^{\otimes (n-i-1)}|1\rangle|x_0,\ldots,x_{i-1}\rangle
\mapsto
\frac{1}{\sqrt{2^{n-i-1}}} |x_0,\ldots,x_{i-1}\rangle|-\rangle \sum_{x_{i+1},\ldots,x_{n-1}=0}^1|x_{i+1},\ldots,x_{n-1}\rangle,
\ee
and
\be
P_{2^n}^\dag:|0\rangle^{\otimes n}
\mapsto \ds\frac{1}{\sqrt{2^n}}\sum_{x_0,\ldots,x_{n-1}=0}^1|x_0,\ldots,x_{n-1}\rangle
\ee
\et

\bo Assume that
\be \label{haar:temp}
P_{2^n}|x_0,\ldots,x_{n-1}\rangle=\lambda_{0,x}|0\rangle^{\otimes n}+\sum_{i=0}^{n-1}(-1)^{x_i}\lambda_{i,x}|0\rangle^{\otimes (n-i-1)}|1\rangle|x_0,\ldots,x_{i-1}\rangle,
\ee
for some parameters $\lambda_{0,x},\lambda_{1,x},\ldots,\lambda_{n-1,x}$. Since $P_{2^n}$ is obtained by normalizing each row of $A_{2^n}$, so $\lambda_{0,x}$ is independent of $x$ and $\lambda_{i,x}$ only depends on $|0\rangle^{\otimes (n-i-1)}|1\rangle|x_0,\ldots,x_{i-1}\rangle$, i.e., $i$ and $|x_0,\ldots,x_{i-1}\rangle$, thus it can be simply denoted as $\lambda_{i,x_0,\ldots,x_{i-1}}$.

For convenience, denote $|x\rangle=|x_0,\ldots,x_{n-1}\rangle$. So we have
\[P_{2^n}=\sum_x \Big(\lambda_0|0\rangle^{\otimes n}+\sum_{i=0}^{n-1}(-1)^{x_i}\lambda_{i,x_0,\ldots,x_{i-1}}|0\rangle^{\otimes (n-i-1)}|1\rangle|x_0,\ldots,x_{i-1}\rangle\Big)\langle x|.\]
Then
\[P_{2^n}^\dag=\sum_x |x\rangle \Big(\lambda_0\langle0|^{\otimes n}+\sum_{i=0}^{n-1}(-1)^{x_i}\lambda_{i,x_0,\ldots,x_{i-1}}\langle0|^{\otimes (n-i-1)}\langle1|\langle x_0,\ldots,x_{i-1}|\Big).\]
Thus
\[P_{2^n}^\dag|0\rangle^{\otimes n} = \sum_x \lambda_0 |x\rangle.\]
Since the norm of $P_{2^n}^\dag|0\rangle^{\otimes n}$ is $1$, so $\lambda_0=1/\sqrt{2^n}$. For any fixed $i$ and $x_0,\ldots,x_{i-1}$,
\begin{eqnarray*}
  P_{2^n}^\dag|0\rangle^{\otimes (n-i-1)}|1\rangle|x_0,\ldots,x_{i-1}\rangle
  &=& \frac{1}{\lambda_{i,x_0,\ldots,x_{i-1}}}\Big(|x_0,\ldots,x_{i-1}\rangle \sum_{x_i,\ldots,x_{n-1}=0}^1(-1)^{x_i}|x_i,\ldots,x_{n-1}\Big) \\
  &=& \frac{\sqrt{2}}{\lambda_{i,x_0,\ldots,x_{i-1}}}\Big(|x_0,\ldots,x_{i-1}\rangle|-\rangle \sum_{x_{i+1},\ldots,x_{n-1}=0}^1|x_{i+1},\ldots,x_{n-1}\Big). \\
\end{eqnarray*}
Also it should has norm 1, so
\[\frac{\sqrt{2}}{\lambda_{i,x_0,\ldots,x_{i-1}}}=\frac{1}{\sqrt{2^{n-i-1}}},\]
thus $\lambda_{i,x_0,\ldots,x_{i-1}}=1/\sqrt{2^{n-i}}$. Substituting the value of $\lambda_0$ and $\lambda_{i,x_0,\ldots,x_{i-1}}$ into (\ref{haar:temp}), we get formula (\ref{haar}).
\hfill $\square$\eo

From (\ref{haar:inverse}), the inverse of quantum Haar  transformation has the following decomposition into elementary unitary transformations
\begin{eqnarray*}
  |0\rangle^{\otimes (n-i-1)}|1\rangle|x_0,\ldots,x_{i-1}\rangle
  &\mapsto& |x_0,\ldots,x_{i-1}\rangle|1\rangle|0\rangle^{\otimes (n-i-1)} \\
  &\mapsto& |x_0,\ldots,x_{i-1}\rangle|-\rangle|0\rangle^{\otimes (n-i-1)} \\
  &\mapsto& |x_0,\ldots,x_{i-1}\rangle|-\rangle\frac{1}{\sqrt{2^{n-i-1}}}\sum_{x_{i+1},\ldots,x_{n-1}=0}^1|x_{i+1},\ldots,x_{n-1}\rangle,
\end{eqnarray*}
where the first step contains $(i+1)(n-i-1)+i=O(n^2)$ swap operations. The second step is a single qubit operation that map $|1\rangle$ to $|-\rangle$. The third step is applying Hadamard transformation on $|0\rangle^{\otimes (n-i-1)}$. So the inverse of quantum Haar  transformation can be implemented in quantum computer with quantum circuit complexity $O(n^2)$. Therefore, we have

\bc
Quantum Haar transformation can be implemented in quantum computer with complexity $O(n^2)$.
\ec

\section{Conclusion}

In this paper, we have considered the problem of constructing efficient quantum unitary transformations. All the unitary transformations constructed in this paper has a explicit formula, which may be easy to use. But it still remains a difficult problem to use new efficient quantum unitary transformations, including the unitary transformations found in \cite{bacon} and \cite{hoyer}, to construct quantum algorithms.

\end{document}